# A Bayesian non-parametric method for clustering high-dimensional binary data

Tapesh Santra[1]
[1]Systems Biology Ireland, University College Dublin, Belfield, Dublin 4, Ireland. Email: tapesh.santra@ucd.ie

## Running title: A non-parametric method for binary data clustering

## Abstract

In many real life problems, objects are described by large number of binary features. For instance, documents are characterized by presence or absence of certain keywords; cancer patients are characterized by presence or absence of certain mutations etc. In such cases, grouping together similar objects/profiles based on such high dimensional binary features is desirable, but challenging. Here, I present a Bayesian non parametric algorithm for clustering high dimensional binary data. It uses a Dirichlet Process (DP) mixture model and simulated annealing to not only cluster binary data, but also find optimal number of clusters in the data. The performance of the algorithm was evaluated and compared with other algorithms using simulated datasets. It outperformed all other clustering methods that were tested in the simulation studies. It was also used to cluster real datasets arising from document analysis, handwritten image analysis and cancer research. It successfully divided a set of documents based on their topics, hand written images based on different styles of writing digits and identified tissue and mutation specificity of chemotherapy treatments.

## Introduction

In many scientific and engineering studies, the presences and absences of a certain set of attributes create binary outcomes which can be used to describe and distinguish individual objects. It is often desirable to group together such objects which have similar features. There are very many methods to achieve such grouping. Some of the most commonly used methods are different versions of k-means [1, 2] and hierarchical clustering [3, 4] algorithms. One of the major drawback of these algorithms is that these require prior knowledge of the number of clusters ($K$). Several methods, such as Gap-statistic [5], Minimum Description Length (MDL) [6] etc., have been developed to be used alongside k-means or hierarchical clustering algorithms in order to estimate the number of clusters ($K$). However, these algorithms have seen limited use in binary data clustering problems. Among other notable efforts are model based clustering methods where the data are clustered using some assumed mixture modelling structure of the data. These methods mostly use finite mixture models which are convex combination of a finite number of simple distributions [7-9]. Bayesian or Akaike information criteria are used with these methods to determine the number of clusters [7-9]. One of the major drawbacks of these methods is the tacit assumption of convex clusters which is seldom realistic. Additionally, these algorithms often use greedy optimization techniques such as Expectation Maximization for selecting optimal cluster labels [7-9]. These methods are prone to converge to local optima, producing sub-optimal results . Full Bayesian methods also exist for model based clustering [10-13]. But many of these methods use Markov Chain Monte Carlo (MCMC) sampling to estimate the posterior distribution of the cluster labels. In addition to the aforementioned issues of model based clustering, these methods also suffer from the *Label Switching Problem* [14] which occur during MCMC sampling and require further additional analysis & processing to obtain some overall form of consensus clustering. This is still very much an open research issue [14-17] with no really consistent solution to this problem available at present. A more systematic approach which avoids many of the problems of model based clustering is Bayesian non-parametric (BNP) methods [18]. Rather than comparing models that vary in complexity (e.g. number of clusters), as is done in case of model based clustering, the BNP approach is to fit a single model that can adapt its complexity to the data. Furthermore, BNP models allow the complexity to grow as more data are observed.  Recently, a few BNP models which are designed to cluster binary data have appeared in literature [19, 20]. These models assume that the data ($\{\boldsymbol{X} = \{X_{ij}, i = 1, \dots, N, j = 1, \dots, D\}$) is generated by a mixture of Bernoulli distributions whose parameters themselves have beta distributions. Furthermore, there is a latent class label ($c_i \in \{1, \dots K \ll N\}$) associated with each observed data ($\boldsymbol{X_i} = \{X_{ij}, j = 1, \dots, D\}$) that indicates which mixture component it belongs to [19, 20]. These latent class labels are thought to be generated by DP and therefore have nonparametric infinite distribution [19, 20]. Current BNP methods assumes that all Bernoulli parameters have the same prior and infer these parameters along with the latent class labels from observed data using variational approaches or MCMC sampling [19, 20].  Here, I present a modified Bayesian Non-parametric Binary Data Clustering algorithm (BNPBDCA) which relies on the same core model as discussed above but differs in some key aspects. For instance, it assumes a different beta prior for each individual feature ($X_{ij}$) of the data, marginalizes the Bernoulli parameters thereby making the posterior distribution of the latent class labels independent of these parameters, and uses simulated annealing to estimate optimal class labels while avoiding the problems associated with variational (prone to converge to local optima [7-9]) and MCMC based methods (suffers from label switching problem [14]). Additionally it enjoys some of the usual benefits of non-parametric method, for instance, it neither makes any parametric assumption about the cluster distributions, nor makes any tacit assumptions of cluster convexity; both of which can be restrictive in some applications [21].

Below I describe the details of BNPBDCA, compare its performance with other BNP and commonly used non Bayesian methods such as kmeans and hierarchical agglomerative clustering using simulated data, implement it on three real datasets involving document analysis, computer vision and cancer research. A MATLAB implementation of BNPBDCA algorithm along with all simulated and real datasets can be freely downloaded from [https://github.com/SBIUCD/BNPBDCA.git](https://github.com/SBIUCD/BNPBDCA.git).

# BNPBDCA, a Bayesian non-parametric algorithm for binary data clustering

## Mathematical formulation

Consider a collection of $N$ objects, each of which is described by a binary feature vector ($\boldsymbol{X_i}$) of length $D$, i.e. $\boldsymbol{X_i} = \{X_{ij}, j = 1 \dots D\}$. Each element ($X_{ij}$) of this vector ($\boldsymbol{X_i}$) can have one of two values *0,1* , i.e. $X_{ij} \in \{0,1\}$. The aim of clustering is to assign a label ($c_i \in \{1, \dots K \ll N\}$) to each of the $N$ objects so that objects with "similar" features are assigned the same label and therefore are gorouped together. In statistical terms, this means grouping together the objects whose feature vectors are samples of the same distribution. Here, the feature vectors ($\boldsymbol{X_i}$) are assumed to have the following prior distributions

$$X_{ij}|c_i = k, p_{jk} \sim Bernoulli(p_{jk}) \quad \text{(a)}$$

$$p_{jk} \sim Beta(a_{jk}, b_{jk}) \quad \text{(b)}$$

$$c_i|\Pi_1, \dots, \Pi_k \sim Discrete(\Pi_1, \Pi_2, \dots, \Pi_k) \quad \text{(c)}$$

$$\Pi_1, \dots, \Pi_k \sim Dirichlet(\frac{\alpha}{K}, \frac{\alpha}{K}, \dots, \frac{\alpha}{K}) \quad \text{(d)} \qquad (1)$$

Eq. 1(a) represents our prior belief that the individual features ($X_{ij}$) of an object ($i$) are independent and for the members of the $k^{th}$ cluster, these can be 1 or 0 with probabilities $p_{jk}$ and $1 - p_{jk}, j = 1, \dots, D$ respectively. Eq. 1(b) indicates that $p_{jk}$ has Beta priors with shape parameters $a_{jk}, b_{jk}$ respectively, where $a_{jk}, b_{jk}$ represent our prior (to observing the data) belief about the frequency with which the $j^{th}$feature occur ($X_{ij} = 1$) in the $k^{th}$ cluster. For instance, if our prior knowledge suggests that the $j^{th}$ feature is relatively sparse in $k^{th}$ cluster one should choose ($b_{jk} \gg a_{jk}$) and vice versa. Eq. 1(c) represents our prior assumption that an object ($i$) belongs to the $k^{th}$cluster with probability $\Pi^k$, i.e. $P(c_i = k) = \Pi_k$ where $\sum_{k=1}^{K} \Pi_k = 1$ . Here, $\Pi_k, k = 1, \dots, K$ are unknown parameters and have Dirichlet distribution with hyper-parameters $\alpha, K$ (Eq. 1(d)). Here $\alpha$ is the concentration parameter and higher values of $\alpha$ allow $\Pi_k$ to decay slower as $k \to \infty$ and thus encourage more clusters. The interdependence between different variables, parameters and hyper-parameters of the model in Eq. 1 is illustrated in Fig.1.

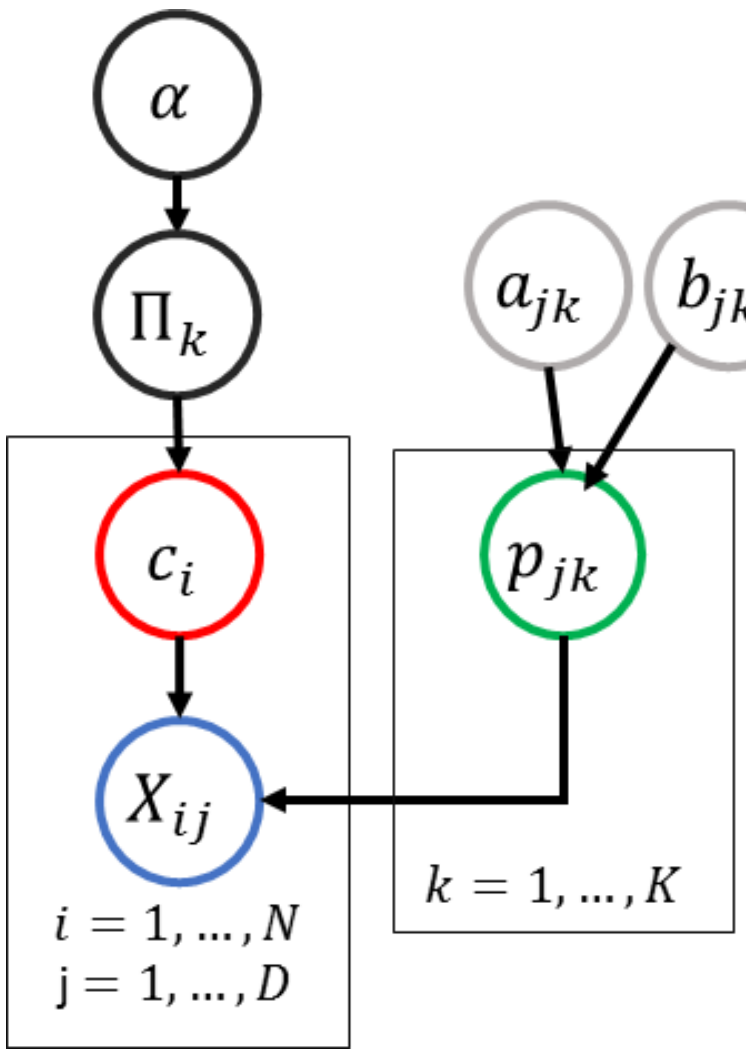

**Figure 1:** Graphical model representation of the Beta-Bernoulli mixture model. $X_{ij}$ are observations which are governed by the Bernoulli distribution parameter $p_{jk}$ which has beta distributions with parameters $a_{jk}, b_{jk}$. The latent variable $c_i$ indicates cluster assignment of $X_{ij}$. $c_i$ takes a value between $\{1, \dots, K\}$ with probabilities $\Pi_1, \dots, \Pi_k$ which has Dirichlet distribution with concentration parameter $\alpha$.

Clustering the binary data ($\boldsymbol{X} = \{\boldsymbol{X_i}: i = 1, \dots, N\}$) amounts to calculating the posterior probability ($P(c_i = k \,|\boldsymbol{c_{-i}}, \boldsymbol{X})$ of allocating the $i^{th}$ object to the $k^{th}$ cluster( $c_i = k$), given the dataset ($\boldsymbol{X} = \{\boldsymbol{X_i}: i = 1, \dots, N\}$) and cluster labels ($\boldsymbol{c_{-i}}$) of the remaining objects. This can be done using Bayes' rule as shown below

$$P(c_i = k, |\boldsymbol{c_{-i}}, \boldsymbol{X}) = \frac{P(\boldsymbol{X_i}|c_i = k, \boldsymbol{c_{-i}}, \boldsymbol{X_{-i}}) \times P(c_i = k|\boldsymbol{c_{-i}})}{\sum_{k=1}^{K} P(\boldsymbol{X_i}|c_i = k, \boldsymbol{c_{-i}}, \boldsymbol{X_{-i}}) \times P(c_i = k|\boldsymbol{c_{-i}})} \quad (4)$$

Where $P(\boldsymbol{X_i}|c_i = k, \boldsymbol{c_{-i}}, \boldsymbol{X_{-i}})$ is the likelihood of object $i$ belonging to cluster $k$, given the cluster labels and binary features of other objects. It can be calculated as follows. We first calculate the likelihood of the members ($\{\boldsymbol{X_j}: c_j = k \;\forall\; j \neq i\}$) of cluster $k$ excluding object *i*.

$$P\left(\boldsymbol{X_l}: c_l = k \;\forall\; l \neq i | p_{jk}, j = 1, \dots, D\right) = \prod_{j=1}^{D} p_{jk}^{N_{-i,jk}} \left(1 - p_{jk}\right)^{N_{-i,k}} \quad (5)$$

Here, $N_{-i,jk} = \sum_{l=1, l \neq i}^{N_{-i,k}} X_{lj}$ . Recall that $p_{jk}$ has a Beta prior (Eq. 1c). Hence, by Bayes' rule the posterior of $p_{jk}$ is

$$P\left(p_{jk}|\boldsymbol{X_{-i}}, \boldsymbol{c_{-i}}\right) = \frac{P\left(\boldsymbol{X_l}: c_l = k \;\forall\; l \neq i | p_{jk}, j = 1, \dots, D\right) \times P\left(p_{jk}\right)}{\int P\left(\boldsymbol{X_l}: c_l = k \;\forall\; l \neq i | p_{jk}, j = 1, \dots, D\right) \times P\left(p_{jk}\right) dp_{jk}} \propto Beta(a_{jk} + N_{-ijk}, b_{jk} + N_{-i,k} - N_{-ijk}) \;. \quad (6)$$

The likelihood term in Eq. 4 can be calculated using Eq. 6 as shown below:

$$\begin{aligned} P(\boldsymbol{X_i}|c_i = k, \boldsymbol{X_{-i}}, c_{-i}) &= \int P\left(\boldsymbol{X_i}|c_i = k, p_{jk}\right) \times P\left(p_{jk}|\boldsymbol{X_{-i}}, \boldsymbol{c_{-i}}\right) dp_{jk} \\ &= \int \prod_{j=1}^{D} p_{jk}^{X_{ij}} \left(1 - p_{jk}\right)^{1 - X_{ij}} \times Beta\left(a_{jk} + N_{-ijk}, b_{jk} + N_{-i,k} - N_{-ijk}\right) dp_{jk} \\ &\propto \frac{B\left(a_{jk} + N_{-ijk} + X_{ij}, b_{jk} + N_{-i,k} - N_{-ijk} + 1 - X_{ij}\right)}{B\left(a_{jk} + N_{-ijk}, b_{jk} + N_{-i,k} - N_{-ijk}\right)} \end{aligned} \quad (7)$$

Furthermore, the *Discrete-Dirichlet* cluster membership model in Eq. 1c,d allows us to calculate the conditional probability ($p(c_i = k|\boldsymbol{c_{-i}})$) that an object (*i*), which is not allocated to a cluster yet, will occupy a certain cluster (*k*), when the remaining objects are already assigned one of the *K* clusters, i.e. the probability of $c_i = k$, given the cluster labels ($\boldsymbol{c_{-i}}$) of the remaining objects. When we have finite number of non-empty clusters (*K*) this conditional is [18-20]

$$p(c_i = k|\boldsymbol{c_{-i}}) = \frac{N_{-i,k} + \alpha/K}{N - 1 + \alpha} \quad (8)$$

Here, $N_{-i,k}$ represents the number of objects in the $k^{th}$ cluster excluding the $i^{th}$ object. However, when the number of cluster in a dataset is unknown, it is assumed that there are an infinite number of clusters available [18-20], and only $K$ of those are occupied by the objects other than the $i^{th}$ object [18-20]. While choosing a cluster label for the $i^{th}$ object we can either choose one of the $K$ non-empty clusters, or decide to allocate it to an empty cluster with the following conditional probabilities [18-20]

$$p(c_i = k|\boldsymbol{c}_{-i}) = \frac{N_{-i,k}}{N-1+\alpha} \quad \text{(a)}$$

$$p\left(c_i \neq c_j \; \forall i \neq j \middle| \boldsymbol{c}_{-i}\right) = \frac{\alpha}{N-1+\alpha} \quad \text{(b)} \quad (9)$$

Replacing Eq. 7 & 9 in Eq. 4 yields

$$P(c_i = k, |\boldsymbol{c}_{-i}, \boldsymbol{X}) \propto N_{-i,k} \frac{B\left(a_{jk}+N_{-ijk}+X_{ij}, b_{jk}+N_{-i,k}-N_{-ijk}+1-X_{ij}\right)}{B\left(a_{jk}+N_{-ijk}, b_{jk}+N_{-i,k}-N_{-ijk}\right)} \quad \text{(a)}$$

$$P\left(c_i \neq c_j \; \forall i \neq j \middle| \boldsymbol{c}_{-i}, \boldsymbol{X}\right) \propto \alpha \frac{B\left(a_{jk}+X_{ij}, b_{jk}+1-X_{ij}\right)}{B\left(a_{jk}, b_{jk}\right)} \quad \text{(b)} \quad (10)$$

Eq. 10 represents the posterior distribution of the cluster labels of an object $i$. Although, in theory, setting up Gibbs sampler to draw samples from the above distribution should solve the clustering problem [18, 19], this is not straightforward due to the aforementioned *label switching* problem. Therefore, here I set up a Gibbs sampler with an annealing schedule, which is designed to converge to a point solution for the cluster labels ($c_i, i = 1 \dots N$) rather than on their posterior distributions (Eq. 10) [22]. For this purpose the posteriors in Eq. 8 are first expressed in terms of Boltzman distribution:

$$P(c_i = k, |\boldsymbol{c}_{-i}, \boldsymbol{X}, T) \propto N_{-i,k} \exp\left(-\frac{-\ln\left(\frac{B\left(a_{jk}+N_{-ijk}+X_{ij}, b_{jk}+N_{-i,k}-N_{-ijk}+1-X_{ij}\right)}{B\left(a_{jk}+N_{-ijk}, b_{jk}+N_{-i,k}-N_{-ijk}\right)}\right)}{T}\right) \quad \text{(a)}$$

$$P\left(c_i \neq c_j \; \forall i \neq j \middle| \boldsymbol{c}_{-i}, \boldsymbol{X}, T\right) \propto \alpha \exp\left(-\frac{-\ln\left(\frac{B\left(a_{jk}+X_{ij}, b_{jk}+1-X_{ij}\right)}{B\left(a_{jk}, b_{jk}\right)}\right)}{T}\right) \quad \text{(b)} \quad (11)$$

A Gibbs sampler is then used to sample from the posteriors in Eq. 11, while cooling down the temperature parameter (T) using an annealing schedule.

The overall algorithm is described below:

## Algorithm

$c_i \leftarrow 1 \leq random\ integer \leq K, i = 1, \dots, N$ //Assign each object to a random cluster

$T \leftarrow T_{init}$ // Initialize temperature parameter

$\lambda \leftarrow 0 < constant < 1$ // Assign a value between 0 and 1 to the scheduling parameter

$B_n \leftarrow 0 < constant < M$ // Assign blocksize

$a_{jk} = constant, \; b_{jk} = constant, j = 1 \dots D, k = 1 \dots K$ // Assign values to hyperparameters

$\alpha = constant$// Set Dirichlet concentration parameter to a constant value

*For n=1,…,M* // Repeat for M iterations

*For* $i = 1, \ldots N$ // Iterate through N Objects

$P_{i,k} \leftarrow P(c_i = k, |\boldsymbol{c_{-i}}, \boldsymbol{X}, T), k = 1, \ldots, K$ // Evaluate Eq. 9a for all existing cluster

$P_{i,new} \leftarrow \ P\left(c_i \neq c_j \ \forall i \neq j \middle| \boldsymbol{c_{-i}}, \boldsymbol{X}, T\right)$ //Evaluate Eq. 9b for a potentially new cluster

$Z \leftarrow P_{i,1} + \cdots + P_{i,K} + P_{i,new}$ // Calculate the normalization constant

$P_i \leftarrow \left\{\frac{P_{i,1}}{Z}, \ldots, \frac{P_{i,K}}{Z}, \frac{P_{i,new}}{Z}\right\}$ // Create the cluster allocation probabilities

$c_i \leftarrow sample(P_i)$ // Sample from the cluster allocation probabilitis

*End for*

*If* $(n\%B_n == 0)$ //Perform cooling after each $B_n$ iterations

$T = T * \lambda$ *;* //Cool down the annealing temperature

*End if*

*End for*

## Choices of parameters and hyper-parameters

To implement the above algorithm one needs to choose the values of the shape parameters $(a_{jk}, b_{jk})$ of the beta distributions, the concentration parameter $(\alpha)$ of the Dirichlet process, the initial temperature $(T_{init})$, the cooling interval $(B_n)$ and the cooling factor $(\lambda)$ of the simulated annealing method. A common practice is to assign some reasonable constant values to the beta and Dirichlet prameters, although, a sampling scheme for the Dirichlet concentration parameter $(\alpha)$ has also been implemented in a full Bayesian implementation of the Beta-Bernoulli Dirichlet scheme [19]. However, sampling or optimization of these parameters on the fly is computationally expensive. Therefore, these parameters are assigned fixed values in this paper. The beta parameters $(a_{jk}, b_{jk})$ represent our prior belief about how many times the $j^{th}$ feature does and does not occur in the $k^{th}$ cluster respectively. In the following implementations $a_{jk}$ is set to 1, and $b_{jk}$ is empirically estimated $b_{jk} = \frac{N}{\sum_{i=1}^{N} x_{ij}}$, to reflect the rather naïve prior belief that most features occur in different clusters with *roughly* the same level of sparsity as in the observed data. The concentration parameter $(\alpha)$ of the Dirichlet distribution determine the number of clusters, with larger values of $\alpha$ leading to more clusters. $\alpha$ was set to 1 to avoid fragmentation of data into too many clusters. The performance of the above algorithm is also dependent on the simulated annealing parameters. Ideally, the initial temperature $(T_{init})$ should be large $T_{init} \geq 1$, the cooling factor $(\lambda)$ should be less than but close to 1, i.e. $\lambda < 1, \lambda \approx 1$, cooling should be performed at each iteration $(B_n = 1)$ and the algorithm should run for a large number of iterations. Such parameter set up is more likely to find the globally optimal cluster structure but is highly computation intensive. Here, $T_{init}$ was set to 1, $\lambda$ was set to 0.9 and $B_n$ was set to 20 for the following implementations.

# Results

## Simulated data

We first implemented the above algorithm on a number of toy datasets with different number of data points (*N*), features (*D*), levels of information (*Sd*) and noise (*Sn*). Each dataset was created by first creating a $N \times D$ matrix of 0s, then randomly dividing its elements in roughly five clusters, randomly selecting *Sd* percentage of features in each cluster to represent information and setting these features to 1, and finally randomly toggling the state of *Sn* percentage of all data points to add spurious noise. A total of 10 datasets were generated (see Tab.1 for details, datasets available from https://github.com/SBIUCD/BNPBDCA.git ), each of which was clustered using four different algorithms, BPNBDCA, dpmm_bernoulli [19] which is a full Bayesian implementation of BNP clustering algorithm (source code originally downloaded from http://tinyurl.com/oly2wqr), MATLAB's proprietary hierarchical agglomerative clustering algorithm, and GAP statistic [5] with k-means algorithm (source code provided with the implementation of BNPBDCA ). The clustering results of each algorithm was then compared with the ground truth by calculating the percentage of data points that were correctly clustered. The performances of all four algorithms on all ten datasets are shown in Tab. 1.

**Table 1:** Comparison of four clustering algorithms on 10 simulated datasets. The first, second and third columns contain the names, sizes and dimensions of the datasets, the fourth and fifth columns contain percentages of information and noise. The sixth-tenth columns contain the accuracies of different clustering algorithms (indicated in the header) in terms of percentage of data-points allocated to the correct cluster.

| Dataset | N | D | Sd | Sn | BNPBDCA | DPMM BERNOULLI | HIERARCHICAL | GAP STATISTICS |
|---|---|---|---|---|---|---|---|---|
| **Dataset1** | 200 | 500 | 10 | 20 | 97.5 | 30 | 31 | 87.5 |
| **Dataset2** | 100 | 500 | 5 | 20 | 100 | 36 | 47 | 59 |
| **Dataset3** | 1000 | 100 | 10 | 10 | 82.3 | 34.3 | 33.9 | 79.9 |
| **Dataset4** | 100 | 1000 | 20 | 30 | 100 | 97 | 74 | 77 |
| **Dataset5** | 200 | 200 | 20 | 20 | 100 | 33.5 | 32 | 77.5 |
| **Dataset 6** | 200 | 200 | 5 | 10 | 98 | 51 | 91 | 74.5 |
| **Dataset 7** | 200 | 200 | 10 | 10 | 100 | 42.5 | 70.5 | 94 |
| **Datset 8** | 200 | 500 | 20 | 20 | 100 | 66 | 51.5 | 94 |
| **Dataset9** | 200 | 500 | 2 | 5 | 99.5 | 49.5 | 53.5 | 83.5 |
| **Dataset10** | 500 | 500 | 20 | 50 | 31.2 | 31.2 | 30.6 | 31.2 |

The above table shows that BNPBDCA performed the best among all four algorithms, clustering nine out of ten datasets with high accuracy and failing only when 50% of data was random noise (Dataset10). The next best performer was Gap-Statistic + kmeans followed by hierarchical clustering. Surprisingly, dpmm_bernoulli which is a full Bayesian implementation of the Beta-Bernoulli-Dirichlet process failed to identify the cluster structure of most datasets except one (dataset 4) and therefore performed the worst. This can be due to many possible reasons, e.g. it assumes that all features ($X_{ij}$) across all clusters have the same beta prior which may be insufficient to model realistic datasets where different features occur at different frequencies. BNPBDCA avoids this problem by assigning different prior for each feature. Additionally, dpmm_bernoulli implements a sampling scheme for the concentration parameter of the Dirichlet distribution which may lead to very small values of this parameter resulting in too few clusters. BNPBDCA avoids this problem by fixing the concentration parameter to a reasonable constant value. Last but not least, dpmm_bernoulli uses MCMC sampling which are known to be inefficient for solving clustering problems, whereas BNPBDCA uses simulated annealing which are designed to find globally optimal clusters.

Encouraged by the above results we used BNPBDCA for analysing three real datasets. Below we describe the results of these analysis.

## BBC sport dataset

One of the most active field of machine learning research is document clustering. Documents are clustered based on presence or frequency of different words in them. However, there can be thousands of unique words in a set of document and therefore the feature-vector that describes a document is very large. Additionally, there are usually relatively few common words between different documents. This results in highly sparse feature matrix (a matrix that represents the presence/frequency of different words in different documents) which poses further difficulty in clustering. To see whether BNPBDCA can be effective in such scenarios I used it to cluster a dataset that contains 737 documents corresponding to sports related news articles published in the BBC Sport website between the years 2004 and 2005 (freely available for download from http://mlg.ucd.ie/datasets/bbc.html ). These documents were previously analysed by Greene et. al. [23] who found 4613 unique words, excluding the common stop words (e.g. a, an, am, see http://mlg.ucd.ie/files/datasets/stopwords.txt for a full list), in these documents. Here, only those words which appeared more than once in an article and found in more than 10 articles were considered for clustering. 406 words met this criteria. For each document, a binary vector was constructed to indicate the presence or absence of these words in that document. This resulted in a $737 \times 406$ binary matrix (Fig. 2A). Each row of this matrix represents an article, each column represents a word and each element represents whether the word in its corresponding column is present in the article in its corresponding row. BNPBDCA divided the matrix into 16 clusters (Fig 2B). Word clouds consisting of the most frequently occurring words in these clusters are shown in Fig. 2C-R. These figures suggest each of these clusters represent different sports related topics. For instance, football related articles are roughly divided into six clusters, containing articles discussing about football clubs, champions league, English league, world cup, match wins and details of a certain game. Cricket related articles are divided into three broad categories, pretext, analysis and general discussion about cricket matches. Athletics related articles are divided into three broad categories, Olympic, world championships, and drug test related discussions. Rubgy and tennis related articles were grouped into two distinct clusters. Injury related articles were grouped in a separate cluster. These results suggest that BNPBDCA can identify different topics among a set of documents and divide them into groups according to their topics.

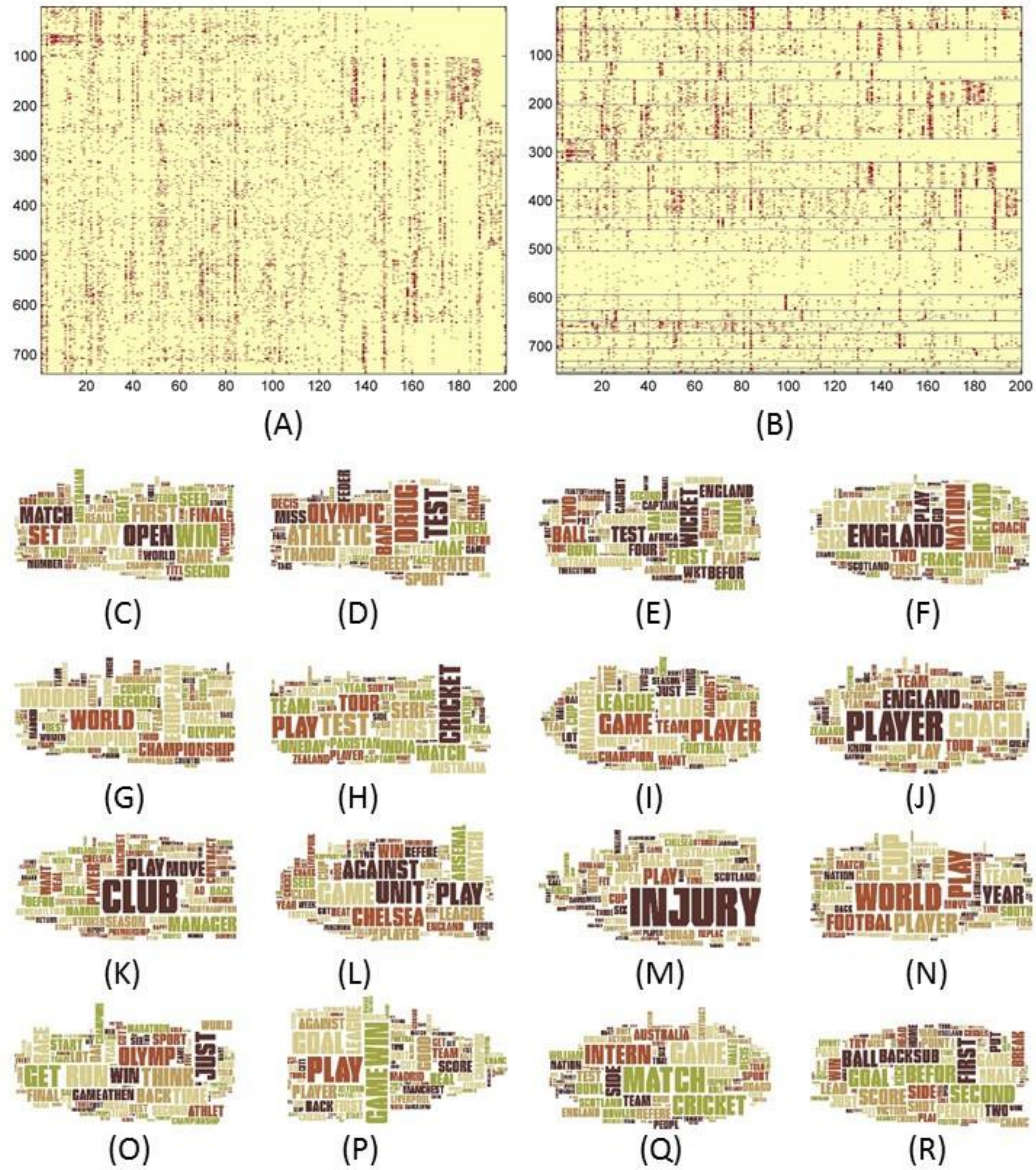

**Figure 2:** (A,B) Raw and clustered binary matrix of the BBC news dataset. Here red and yellow represents 1 and 0 respectively, X and Y axis represent feature dimension and data points (documents) respectively. (C-R) Word clouds showing the most frequently occurring words in different clusters.

## MNIST handwritten digits dataset

BNPBDCA was tested on part of the MNIST dataset (see http://yann.lecun.com/exdb/mnist/ for details) which contain 60000 grayscale images of handwritten digits 0-9 . Each image has $20 \times 20 = 400$ pixel features which were converted to binary by setting all pixels with values greater than 128 to 1 and the remaining pixels to 0.

Only first one thousand images were used for clustering. This resulted in a $1000 \times 400$ binary matrix (Fig. 3A) which was then clustered using BNPBDCA. It identified 36 clusters (Fig 3B), most of which represent different styles of writing digits (Fig 3C). Some clusters also contain noisy/unclear images, and in some cases where certain writing styles of two different digits (e.g. 3 and 5) closely resemble each other such images were also grouped in one cluster (instead of two clusters). Overall, BNPBDCA was not only able to identify, different styles of writing digits, it could also identify bad/noisy/corrupted images and grouped those into separate clusters.

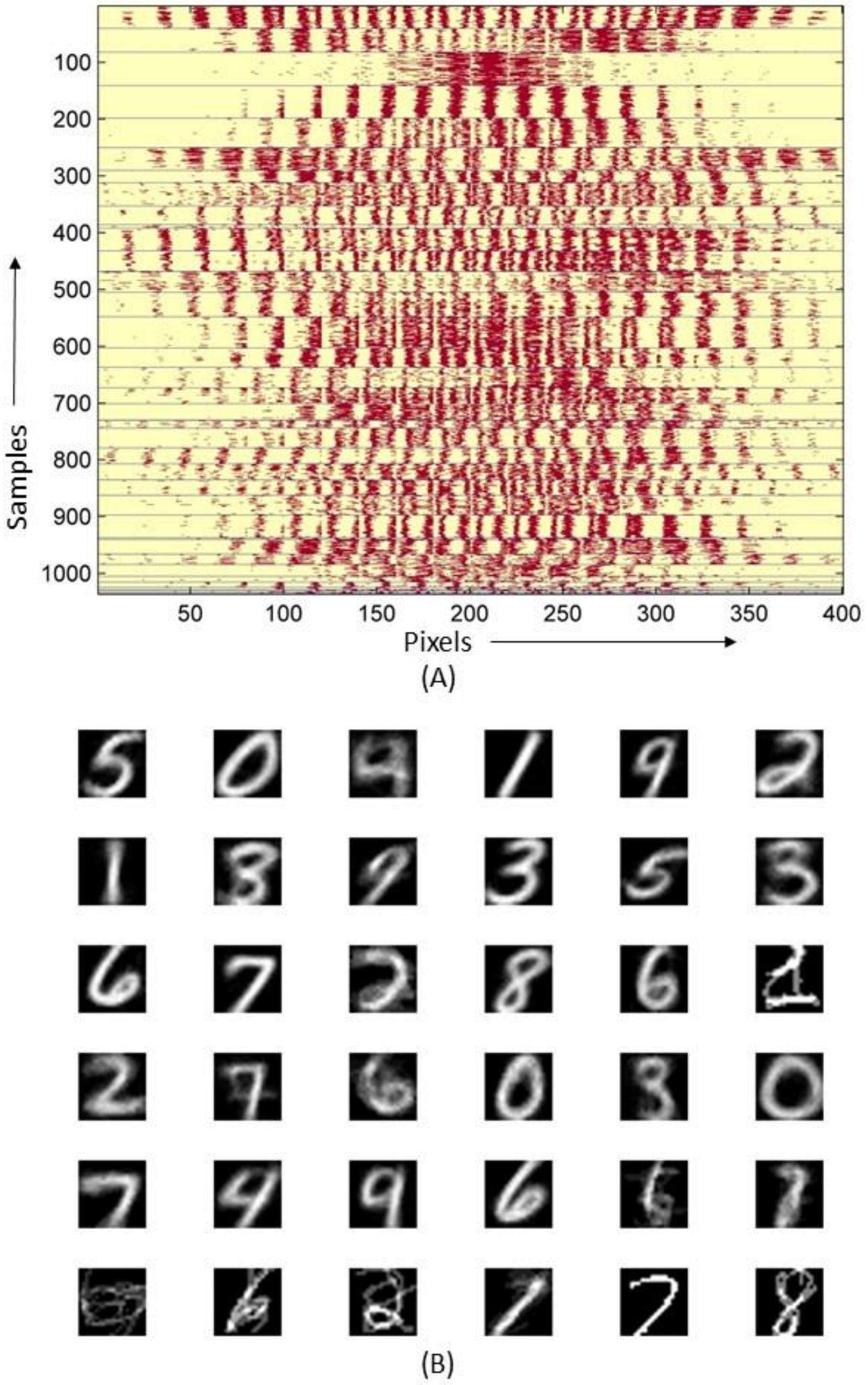

**Figure 3:** (A) Clustered binary matrix derived from MNIST dataset. (B) Average image of each cluster.

## Cancer drug response data

Chemotherapeutic drugs are some of the most common form of treatments for cancer patients [24]. Unfortunately, these drugs do not work on all patients [25]. Currently there are no consensus on identifying patients who will or will not respond to a certain chemotherapy drug [25]. Currently there is a concerted effort to understand tissue/mutation specificity of different therapeutic options for cancer patients. As part of this effort, Yang et. al. performed a large number of experiments where they measured the responses of 708 different types of cancer cells to 99 approved and experimental cancer drugs. Here, I focus on six chemotherapy drugs (Doxorubicine, Etoposide, Gemcitanib, Cisplatin, Docetaxel, Methotrexate) which are commonly used in clinics to treat advance stage cancers. The cells are characterized by the organ/tissue they came from (e.g. lung, breast, blood etc.) and the presence/absence of a number of commonly occurring cancer mutations. The drug responses are given in terms of IC50 values which are commonly used by pharmacologists to indicate the potency of a drug, where smaller IC50 values indicate higher potency. The dataset was first transformed into binary data by (a) creating binary indicator variables for each tissue/organ type, that indicate whether a cell is from a certain tissue/organ or not, (b) using binary mutations indicators to indicate whether a cell contains certain mutation or not, and (c) thresholding the IC50 values of each drug to its lowest $20^{th}$ percentile (of the IC50 values of the same drug across all cell types) and setting all values below this threshold to 1 and remaining values to 0. The dataset contains a large number of missing data-points, e.g. mutational status of some genes and responses to certain drugs are not available for all cells types. All such cases were removed. Two mutations (TP53 and CDK2NA) occurred in almost all cells and were also removed from the data, leaving eight mutational status (KRAS, BRAF, PIK3CA, SMAD4, RB1, PTEN, ERBB2, MYC) to characterize each cell. Cells which had none of these mutations were also removed from further analysis. The resulting dataset is a $298 \times 27$ matrix, containing binary responses of 298 cells from 13 different organs/tissues, containing at least one of the eight aforementioned mutations, to six different chemotherapeutic drugs. BNPBDCA identified thirteen clusters in the data (Fig. 4A). The frequency of different features in each cluster are plotted in Fig. 4B-N. These figures reveal that five of the six drugs (all except Methotrexate) are most effective on tumours which predominantly occur in nervous system, bone, urogenital system and digestive system, and has at-least one of PI3KCA, RB1, PTEN mutations (Fig. 4E). The tumours that grow in pancreas and aero-digestive system and has at least one of KRAS, PI3KCA, SMAD4 mutations are treated most effectively by Docetaxel. Doxorubicin also saw moderate success on these cells, but the other drugs (Etoposide, Gemcitanib, Cisplatin, Methotrexate) had limited effect (Fig. 4K). Methotrexate was most effective on blood cancer cells with predominantly PTEN mutation, while the remaining drugs had limited effect on these cells (Fig. 4H). Methotrexate also had moderate success on digestive system tumours with MYC mutation (Fig. 4J). None of the chemotherapy drugs had significant effects on lung, breast and skin cancer cells (Fig. 4 B,C,D,G,M). Cancer cells with KRAS, BRAF and ERBB2 mutations were also not significantly affected by these drugs (Fig. 4 C, G,N). The above results suggest that BNPBDCA identified tissue and mutation specificity of different chemotherapy drugs. It also revealed that some of the mutations (KRAS, BRAF, ERBB2) which are commonly used as predictors of several cancer drugs [26] may not be reliable predictors of chemotherapy treatments.

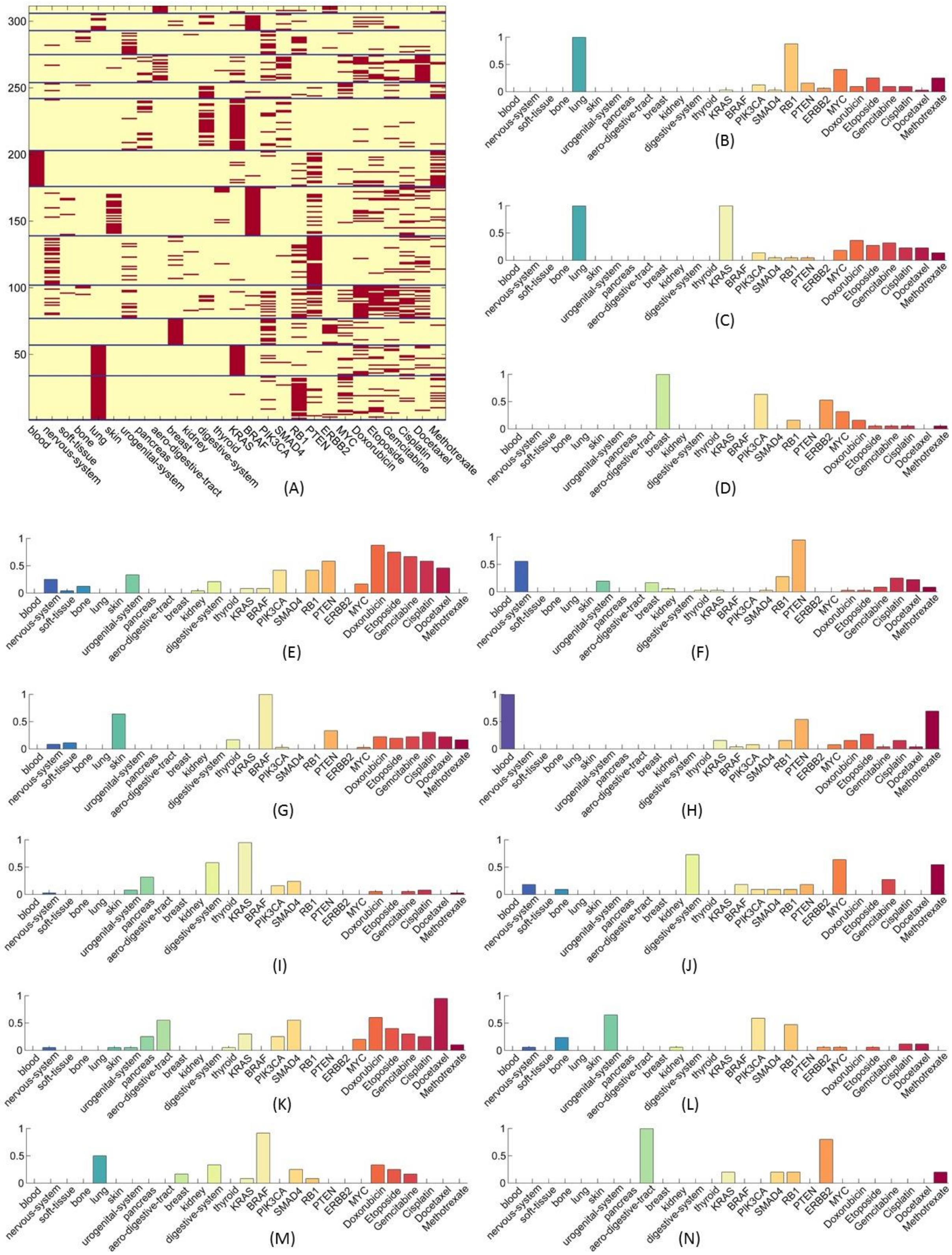

Figure 4: (A) Clustered binary data derived from the cancer dataset. (B-N) frequencies of different features in each cluster.

# Discussion

The explosive growth in data generation and storage capabilities have outpaced our abilities to interpret and organize these data in a meaningful way. These data usually contains rich description of events or objects which require highly sophisticated tools for meaningful analysis. However, in some cases a broad understanding of the data is favoured over detailed analysis. For instance grouping together documents of similar topics, images of similar shapes, genes or proteins of similar functional annotations, patients with similar clinical profile, persons with similar interests etc. may be preferred in some applications over reconstructing a more detailed picture of interdependence between these objects. In many cases representing such data in binary form and the clustering them into groups serves the purpose. Yet, the area of clustering binary data has been largely ignored by mainstream machine learning researchers. In this paper, we present a fully unsupervised binary data clustering algorithm which can not only cluster binary data, but also identify the number of clusters in a dataset. The presented method uses non-parametric Bayesian formulation along with simulated annealing for clustering binary data. It solves many problems of existing methods, e.g. it can estimate the number of clusters in a datasets and it is more likely to provide a globally optimal cluster structure than other similar methods. It outperformed a similar algorithm and two other commonly used clustering algorithms in a simulation study. It was then used to analyse binary data originating from document analysis, image analysis and cancer research and it was able to successfully group together documents with similar topics, images with similar shapes, and cancer cells with similar features. Therefore, the method proposed in this paper can be a useful tool for data analysis in different fields of science and technology.

However, the performance of the above method depends on the values of its hyper-parameters and the annealing schedule of the simulated annealing method. In our paper, values of these parameters are either estimated empirically or given a reasonable constant value based on common practice in literature. A more systematic approach should be adopted to estimate optimal values of these parameters. Previous attempts, where some of these parameters were sampled within the clustering algorithm, did not work well in our simulation study [19], suggesting the need for alternative methods. Furthermore, designing an optimal annealing schedule for simulated annealing based methods is still an open area of research [27]. These will be the primary focus of my future research in this area.

# References

1. Likas, A., N. Vlassis, and J.J. Verbeek, *The global k-means clustering algorithm.* Pattern recognition, 2003. **36**(2): p. 451-461.
2. Ordonez, C. *Clustering binary data streams with K-means*. in *Proceedings of the 8th ACM SIGMOD workshop on Research issues in data mining and knowledge discovery*. 2003. ACM.
3. Tamasauskas, D., V. Sakalauskas, and D. Kriksciuniene. *Evaluation framework of hierarchical clustering methods for binary data*. in *Hybrid Intelligent Systems (HIS), 2012 12th International Conference on*. 2012.
4. Hands, S. and B. Everitt, *A Monte Carlo Study of the Recovery of Cluster Structure in Binary Data by Hierarchical Clustering Techniques.* Multivariate Behavioral Research, 1987. **22**(2): p. 235-243.

5. Tibshirani, R., G. Walther, and T. Hastie, *Estimating the number of clusters in a data set via the gap statistic.* Journal of the Royal Statistical Society: Series B (Statistical Methodology), 2001. **63**(2): p. 411-423.
6. González-Rubio, J. and F. Casacuberta, *Improving the Minimum Description Length Inference of Phrase-Based Translation Models*, in *Pattern Recognition and Image Analysis*. 2015, Springer. p. 219-227.
7. Cagnone, S. and C. Viroli, *A factor mixture analysis model for multivariate binary data.* Statistical Modelling, 2012. **12**(3): p. 257-277.
8. Gollini, I. and T.B. Murphy, *Mixture of latent trait analyzers for model-based clustering of categorical data.* Statistics and Computing, 2014. **24**(4): p. 569-588.
9. Tang, Y., R.P. Browne, and P.D. McNicholas, *Model based clustering of high-dimensional binary data.* Computational Statistics & Data Analysis, 2015. **87**: p. 84-101.
10. Breese, J.S., D. Heckerman, and C. Kadie. *Empirical analysis of predictive algorithms for collaborative filtering*. in *Proceedings of the Fourteenth conference on Uncertainty in artificial intelligence*. 1998. Morgan Kaufmann Publishers Inc.
11. Dunson, D.B., *Bayesian latent variable models for clustered mixed outcomes.* Journal of the Royal Statistical Society. Series B, Statistical Methodology, 2000: p. 355-366.
12. Sutskever, I., J.B. Tenenbaum, and R.R. Salakhutdinov. *Modelling relational data using Bayesian clustered tensor factorization*. in *Advances in neural information processing systems*. 2009.
13. Turner, R.M., R.Z. Omar, and S.G. Thompson, *Bayesian methods of analysis for cluster randomized trials with binary outcome data.* Statistics in medicine, 2001. **20**(3): p. 453-472.
14. Stephens, M., *Dealing with label switching in mixture models.* Journal of the Royal Statistical Society: Series B (Statistical Methodology), 2000. **62**(4): p. 795-809.
15. Li, H. and X. Fan, *A pivotal allocation based algorithm for solving the label switching problem in Bayesian mixture models.* Journal of Computational and Graphical Statistics, 2014(just-accepted): p. 00-00.
16. Papastamoulis, P., *Handling the label switching problem in latent class models via the ECR algorithm.* Communications in Statistics-Simulation and Computation, 2014. **43**(4): p. 913-927.
17. Rodríguez, C.E. and S.G. Walker, *Label switching in bayesian mixture models: Deterministic relabeling strategies.* Journal of Computational and Graphical Statistics, 2014. **23**(1): p. 25-45.
18. Hjort, N., C. Holmes, and P. Mueller, *SW (2010). Bayesian Nonparametrics: Principles and Practice*. Cambridge University Press.
19. Gershman, S.J. and D.M. Blei, *A tutorial on Bayesian nonparametric models.* Journal of Mathematical Psychology, 2012. **56**(1): p. 1-12.
20. Martin, A.P., et al., *Lapatinib Resistance in HCT116 Cells Is Mediated by Elevated MCL-1 Expression and Decreased BAK Activation and Not by ERBB Receptor Kinase Mutation.* Molecular Pharmacology, 2008. **74**(3): p. 807-822.
21. Li, J., S. Ray, and B.G. Lindsay, *A Nonparametric Statistical Approach to Clustering via Mode Identification.* Journal of Machine Learning Research, 2007. **8**(8): p. 1687-1723.
22. Ingber, L., *Simulated annealing: Practice versus theory.* Mathematical and computer modelling, 1993. **18**(11): p. 29-57.
23. Greene, D. and P. Cunningham. *Practical solutions to the problem of diagonal dominance in kernel document clustering*. in *Proceedings of the 23rd international conference on Machine learning*. 2006. ACM.
24. Kadakia, K.C., et al., *Palliative communications: addressing chemotherapy in patients with advanced cancer.* Annals of Oncology, 2012. **23**(suppl 3): p. 29-32.
25. Liang, X.-J., et al., *Circumventing Tumor Resistance to Chemotherapy by Nanotechnology.* Methods in molecular biology (Clifton, N.J.), 2010. **596**: p. 467-488.

26. De Roock, W., et al., *Effects of KRAS, BRAF, NRAS, and PIK3CA mutations on the efficacy of cetuximab plus chemotherapy in chemotherapy-refractory metastatic colorectal cancer: a retrospective consortium analysis.* The lancet oncology, 2010. **11**(8): p. 753-762.
27. Ben-Ameur, W., *Computing the initial temperature of simulated annealing.* Computational Optimization and Applications, 2004. **29**(3): p. 369-385.